\documentclass[reprint,twocolumn,amsmath,amssymb,aps,superscriptaddress]{revtex4-2}

\usepackage{graphicx}% Include figure files
\usepackage{dcolumn}% Align table columns on decimal point
\usepackage{bm}% bold math
\usepackage{graphicx}
\usepackage{amssymb}
\usepackage{epstopdf, epsfig}
\usepackage{yfonts}
\usepackage[colorlinks=true, linkcolor=blue, citecolor=blue, urlcolor=blue]{hyperref}
\usepackage{fontenc}
\usepackage{upgreek}
\usepackage{mathrsfs}
\usepackage{booktabs}
\usepackage{pifont}
\usepackage{amsmath}
\usepackage{bm}
\usepackage{wasysym}
\usepackage{caption}
\usepackage{subcaption}
\usepackage{euscript}
\usepackage{natbib}
\usepackage{lipsum}
\usepackage{xcolor}

\usepackage{orcidlink}

\usepackage{subcaption}%this section justifies revTeX captions
\usepackage{ragged2e}
\DeclareCaptionJustification{justified}{\justifying}
\newcommand{\pd}[2]{\frac{\partial #1}{\partial #2}}

\newcommand{\x}[1]{\textcolor{black}{#1}}
\newcommand{\xx}[1]{\textcolor{black}{#1}}
\newcommand{\zz}[1]{\textcolor{black}{#1}}

\newcommand{\jfm}[1]{\textcolor{blue}{#1}}

\begin{document}

\title{Nonlinear Wave Transformation over Steep Breakwaters}

\author{Saulo Mendes\,\orcidlink{0000-0003-2395-781X}}
\email{saulo.dasilvamendes@unige.ch}
\affiliation{Group of Applied Physics, University of Geneva, Rue de l'\'{E}cole de M\'{e}decine 20, 1205 Geneva, Switzerland}
\affiliation{Institute for Environmental Sciences, University of Geneva, Boulevard Carl-Vogt 66, 1205 Geneva, Switzerland}

\begin{abstract}
Wave shoaling of water waves over mild bottom slopes is well described by linearized theories. However, the analytical treatment of nonlinear wave shoaling subject to rapidly varying bottoms has proven to be elusive in the past decades. As the spatial evolution of the exceedance probability of irregular waves is affected by second-order effects in steepness, the nonlinear shoaling coefficient throughout a symmetrical and steep breakwater is investigated through a stochastic framework. By inverting the effect of slope on normalized wave height distribution, it is possible to obtain a closed-form slope dependence of the nonlinear shoaling coefficient compatible with experiments over steep breakwaters.
\end{abstract}


\maketitle

\section{INTRODUCTION}

Wave characteristics are modified when propagating nearshore until they eventually break. Their transformation due to shoaling is one of the most fundamental coastal processes and has been known since the works of \citet{Green1838} and \citet{Burnside1915}. \xx{Above all, the transformation of wave heights is an essential and required factor for many
coastal and ocean engineering applications. On the other hand, a precise knowledge of the steepness evolution over a shoal is also fundamental, as it affects the solutions of perturbative wave theories. In fact,} primary water wave variables such as group speed \citep{Stokes1847}, dispersion relation \citep{Holthuijsen2007}, pressure underneath waves \citep{Dalrymple984}, energy flux and total energy \citep{Higgins1975b,Ma2020} as well as direct consequences of wave conservation principles such as set-down and set-up \citep{Bowen1968}, wave run-up \citep{daSilva2020}, wave breaking location \citep{Svendsen2005}, longshore and rip currents \citep{Bowen1969b,Bowen1969} are all \xx{affected by the growth in steepness}. For waves approaching the coast, these primary variables can only be predicted accurately if a closed-form for the shoaling coefficient is known. \xx{Although the integral properties of} coastal processes were generalized through the theory of radiation stress \citep{Higgins1962,Higgins1964}\xx{, the} study \xx{of the} amplification of wave height due to a shoal \xx{has been} commonly reduced to the conservation of the energy flux in the absence of refraction, reflection or any \xx{form} dissipation \xx{since \citet{Burnside1915}}. 

Shoaling models include but are not limited to linear theory \citep{Airy1845}, higher-order nonlinear theory of \citet{Stokes1847} as well as cnoidal \citep{Korteweg1895} or \citet{Cokelet1977} theories. Although several \xx{nonlinear} models for wave deformation and transformation for waves encountering bathymetric changes exist, these theories do not provide simple closed-form expressions in terms of initial steepness, relative water depth or bottom slope \citep{Eagleson1956,Walker1983,Goda1997}\xx{, justifying the existence of several empirical models \citep{Ratta2018}}. \xx{Naturally,} linear wave theory significantly underpredicts shoaling coefficients for high values of the Ursell number, while nonlinear theories \xx{and empirical formulae thereof do} not capture \xx{the effect of the} slope \xx{magnitude $|\nabla h|$} \citep{Ratta2018}. \xx{Nevertheless, \citet{Iwagaki1973} computed the effect of the shoaling slope magnitude on the surface elevation of cnoidal waves under the assumption that $k_{p}h < \pi/10$ and $|\nabla h| < 1/10$. However, it did not clarify how to compute the wave height from it. Furthermore, the nonlinear correction to the surface elevation is proportional to $|\nabla h|^{-1}$, which instead of recovering the \citet{Green1838} law for the adiabatic case ($|\nabla h| \ll 1/10$) expects wave shoaling at steep slopes to recover this law, in clear contradiction to experimental observation \citep{Walker1983}. In addition, \citet{Iwagaki1973} model has not been formulated for the regions atop a shoal and over the de-shoal, such that it can not be readily extended to a breakwater.} \xx{Indeed, it is widely agreed that no general theory or closed formula assessing the effect of slope magnitude exists for the nonlinear wave transformation, in particular for steep slopes} \citep{LeMehaute1980,Walker1983,Tsai2005,Gupta2017,Srineash2018}. For instance, theoretical and empirical models for the breaker height have been tested and seem to work well for mild slopes ($|\nabla h| < 1/15$) only  \citep{Ratta2000}. 

In addition to the wave transformation of ordinary waves nearshore, the study of rogue waves in this zone has grown exponentially in recent years. The threat to ocean vessels \xx{and structures} posed by rogue waves have recently been discovered to be also important in transitional depths \xx{subject to shoaling} \citep{Trulsen2012,Bolles2019,Chabchoub2019,Trulsen2020,Benoit2021,Mori2023}. The latter phenomenon has been theoretically examined and attributed to among other factors the evolution of the steepness \xx{\citep{Adcock2021c,Mendes2021b,Chabchoub2023}}. However, in order to become predictive, current theories of second-order rogue wave shoaling depend on precise modelling of the \xx{wave transformation} over arbitrary \xx{bathymetry}. \xx{So far, a one-way street has been established between deterministic fundamental properties and the stochastic analysis of hydrodynamics: for a given water wave solution, a unique probability distribution can be obtained.} In the present work, \xx{the stochastic analysis of irregular wave motion is shown to be synchronized to the fundamental property of wave transformation in such a way that is possible to compute the nonlinear shoaling coefficient with more precision than through deterministic (fundamental) methods.} A new theory for the nonlinear \xx{and small amplitude} wave transformation is developed for steep arbitrary slopes over a shoal or de-shoal, thus applicable to beaches and breakwaters. In particular, it is shown how the knowledge of the wave statistics of seas with high rogue wave occurrence is relevant for the computation of the slope-dependent nonlinear shoaling coefficient. Good agreement is found when comparing the theory with the experiments of \citet{Raustol2014}. 

\section{Nonlinear Shoaling:\\ State of the Art}

\xx{Whichever water wave solution is applied to the problem of shoaling, the theoretical evolution of either wave height or wave steepness is invariably performed through the conservation of the energy flux, see \citet{Ratta2018} and \citet{LeMehaute1980} for a review.} To compute the energy flux with the mean water level as \textit{datum}, I adopt the formulation of \citet{Higgins1975b} and \citet{Klopman1990} and restrict it for small amplitude waves $(\zeta \ll h)$:
\begin{eqnarray}
F = \left\langle \int_{-h}^{0} \Big[ p + \frac{1}{2} \rho (u^{2}+w^{2}) + \rho g ( z - \langle \zeta \rangle )    \Big]\, u \, dz  \right\rangle \,\, , 
\end{eqnarray}
\xx{which can be shown to be equivalent to the mean energy level as \textit{datum} subtracted by the mean water level counterpart \citep{Klopman1990,Jonsson1995},}
\begin{eqnarray}
F \equiv \left\langle \int_{-h}^{0} \Big[ \rho g \zeta^{\star} + \frac{1}{2} \rho (u^{2}+w^{2})    \Big]\, u \,  dz  \right\rangle - \rho g  \langle \zeta \rangle c_{p} h     \,\, , 
\label{eq:energyflux}
\end{eqnarray}
where $p$ is the water column pressure in the presence of waves, $(u,w)$ are the horizontal and vertical components of the velocity vector, $c_{p}$ is the phase velocity, $h$ the water depth, $\zeta^{\star}$ the surface elevation corrected by depth function within $\partial \Phi / \partial t$ and $\langle \zeta \rangle$ the mean water level obtained from the momentum balance \citep{Higgins1962} computed by a time average $\langle \cdot \rangle$ operator.

\subsection{Adiabatic Shoaling: Gently Sloping Beaches}

\xx{Typically, gently sloping beaches (adiabatic process) are assumed and no slope magnitude effect is considered for shoaling \citep{Stiassnie1980}, even when the regime of cnoidal waves is reached \citep{Iwagaki1968,Svendsen1972,Shuto1974,Iwagaki1982,Isobe1985}. Such approximation became common practice due to ease in treating shoaling in the absence of reflection or wave deformation \citep{LeMehaute1980}. In this section, I briefly review how the shoaling coefficient is computed through the conservation of energy flux, such that verification of classical formulae for the adiabatic case can be readily generalized to a finite and arbitrary slope magnitude.}

For linear waves, the surface elevation and velocity components for waves traveling over a mild slope are written as \citep{Airy1845,Dalrymple984,Dingemans1997}:
\begin{eqnarray}
\nonumber
\Phi &=& \frac{a\omega}{k} \frac{\cosh{\theta} }{\sinh{\Lambda}}  \sin{\phi} \,\, ; \,\, \zeta^{\star} = a \frac{\cosh{\theta} }{\cosh{\Lambda}} \cos{\phi} \,\, ;
\\
u &=& a\omega  \frac{\cosh{\theta} }{\sinh{\Lambda}}  \cos{\phi} \,\, ; \,\, w =  a \omega \frac{\sinh{\theta} }{\sinh{\Lambda}}  \sin{\phi} \,\, .
\end{eqnarray}
with notation $\theta = k (z+h)$, $\Lambda = kh$ and $\phi = kx - \omega t$ for regular waves. Thus, the energy flux with mean energy level as \textit{datum} (i.e. neglecting the change in mean water level) \x{under an adiabatic shoaling process} reads:
\begin{figure*}
\centering
    \includegraphics[scale=0.6]{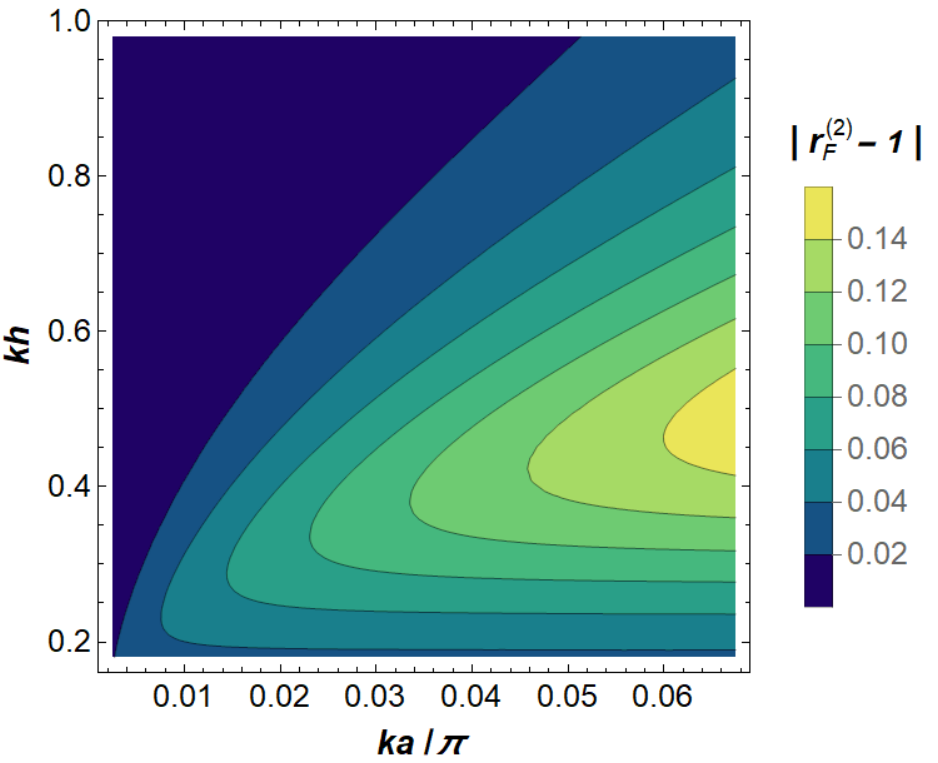}
\caption{Contour plot of the function $r_{F}^{(2)} = F^{(2)}/E^{(2)} c_{g} - 1$, describing the difference in the growth of energy and its flux due to the steepening of second-order waves at an arbitrary depth.}
\label{fig:ratioflux}
\end{figure*}
\begin{eqnarray}
F_{0}^{(1)} &=& \rho g  \int_{-h}^{0} \left\langle \zeta^{\star} u + \frac{u}{2g}  (u^{2}+w^{2})   \right\rangle \,  dz        \quad . 
\end{eqnarray}
Because the terms $u(u^{2}+w^{2}) $ entail odd powers of trigonometric functions such as $\cos^{3}{\phi}$ \zz{and $\sin^{2}{\phi}\cos{\phi}$}, periodicity of these functions imply in $ \langle u(u^{2}+w^{2}) \rangle = 0$. Therefore, \zz{I} arrive at: 
\begin{eqnarray}
\nonumber
F_{0}^{(1)} &=& \rho g  \int_{-h}^{0} \left\langle \zeta^{\star} u   \right\rangle \,  dz  
\\
\nonumber
&=& \rho g a^{2} \cdot   \frac{2\omega }{\sinh{(2\Lambda)}}  \int_{-h}^{0} \cosh^{2}{\theta} \langle \cos^{2}{\phi} \rangle dz    \quad ,
\\
\nonumber
&=&  \rho g a^{2} \cdot   \frac{\omega }{\sinh{(2\Lambda)}}  \int_{-h}^{0} \cosh^{2}{\theta} dz 
\\
\nonumber
&=&  \rho g a^{2} \cdot   \frac{\omega }{\sinh{(2\Lambda)}} \left[ \frac{h}{2}  +  \frac{\sinh{(2\Lambda)}}{4k}  \right] \quad ,
\\
\nonumber
&=&  \rho g a^{2} \cdot   \frac{\omega }{\sinh{(2\Lambda)}} \cdot \frac{\sinh{(2\Lambda)}}{4k}  \left[ 1 + \frac{2\Lambda}{\sinh{(2\Lambda)}}  \right] \,\, , 
\\
\nonumber
&=& \left(  \frac{1}{2} \rho g a^{2} \right) \cdot    \frac{\omega }{2k}  \left[ 1 + \frac{2\Lambda}{\sinh{(2\Lambda)}}  \right] \,\, ,
\\
&=& E^{(1)} \cdot \frac{c_{p}}{2} \left[ 1 + \frac{2\Lambda}{\sinh{(2\Lambda)}}  \right] \equiv E^{(1)} \cdot c_{g}  \quad ,
\label{eq:flux1}
\end{eqnarray}
which is the well-known textbook form for the energy flux. Therefore, the shoaling coefficient for the wave height reads \citep{Svendsen2005,Sorensen2006}:
\begin{equation}
\nabla \cdot (E^{(1)}c_{g} \hat{x}) = 0 \,\, \therefore \,\, K_{s} = \frac{H}{H_{0}} = \sqrt{\frac{c_{g \, 0}}{c_{g}}}   \,\,\, .
\label{eq:Ks}
\end{equation}
As the wavelength decreases due to shoaling, the transformation of the regular wave steepness in linear theory can be computed \citep{Eagleson1956,Svendsen2005}:
\begin{equation}
K_{\varepsilon} =  \frac{\varepsilon}{\varepsilon_{0}} \equiv \frac{H}{H_{0}} \frac{\lambda_{0}}{\lambda}  = \frac{1}{\tanh{kh}} \left[ \frac{2\cosh^{2}{kh}}{2kh + \sinh{2kh}} \right]^{1/2}  \quad .
\label{eq:varepsilons2}
\end{equation}
The derivation of eq.~(\ref{eq:flux1}) is notably simpler than the one provided by \citet{Higgins1975b}. The advantage of the above direct approach is that it does not depend on relationships between integral properties such as momentum flux, energy flux, mass flux and so on. For waves that become steeper over a steep slope, these relationships have to be carefully revised, while the computation in eq.~(\ref{eq:flux1}) concerns only its own dependence on the inhomogeneity of the wave evolution.

\subsubsection{Secon-order Waves Travelling over a Shoal}

Second-order regular waves has a potential \citep{Dingemans1997}:
\begin{equation}
\Phi = \frac{a\omega}{k} \frac{\cosh{\theta} }{\sinh{\Lambda}} \sin{\phi} + \left(  \frac{3ka}{8}  \right) \frac{a\omega}{k} \frac{\cosh{(2\theta)} }{\sinh^{4}{\Lambda}} \sin{(2\phi)} \,\, .
\end{equation}
In addition, the pressure field induced by a progressive wave field is \citep{Dalrymple984}:
\begin{eqnarray}
\nonumber
p &=& \rho g (\zeta^{\star}- z) = \rho \frac{\partial \Phi}{\partial t} -   \rho g z \quad ,
\\
\nonumber
\rho \frac{\partial \Phi}{\partial t} &=& \rho g a \Bigg[  \frac{\cosh{\theta} }{\cosh{\Lambda}} \,  \cos{\phi}  
\\
&+& \left(  \frac{3ka}{4}  \right) \frac{\cosh{(2\theta)} }{\cosh{\Lambda} \sinh^{3}{\Lambda}} \, \cos{(2\phi)}    \Bigg] \quad ,
\end{eqnarray}
Similarly, one can easily obtain the second-order horizontal velocity component:
\begin{eqnarray}
u = a\omega \left\{ \frac{\cosh{\theta} }{\sinh{\Lambda}} \cos{\phi}  + \left(  \frac{3ka}{4}  \right) \frac{\cosh{(2\theta)} }{\sinh^{4}{\Lambda}} \cos{(2\phi)} \right\} \,\, .
\end{eqnarray}
Hence, denoting $(\delta u , \delta \zeta^{\star})$ as the second-order additional terms to the linear theory, I can write:
\begin{equation}
F_{0}^{(2)} = \rho g  \int_{-h}^{0} \left\langle \zeta^{\star} \cdot u +  \delta \zeta^{\star} \cdot u + \delta u \cdot \zeta^{\star} + \delta u \cdot \delta \zeta^{\star}    \right\rangle \,  dz \,\, .
\end{equation}
The terms $(\delta u \cdot \zeta^{\star},\delta \zeta^{\star} \cdot u)$ have major integrands $\cos{\phi} \cos{(2\phi)}$ and therefore their averages vanish. Accordingly, the computation leads to:
\begin{figure*}
\minipage{0.47\textwidth}
    \includegraphics[scale=0.54]{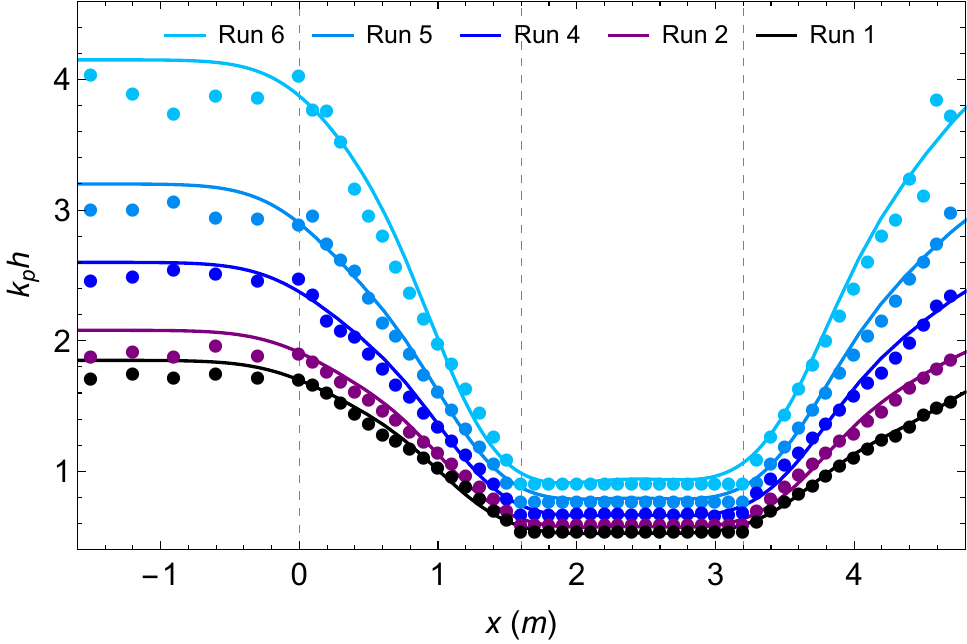}
\endminipage
\hfill
\minipage{0.49\textwidth}
    \includegraphics[scale=0.54]{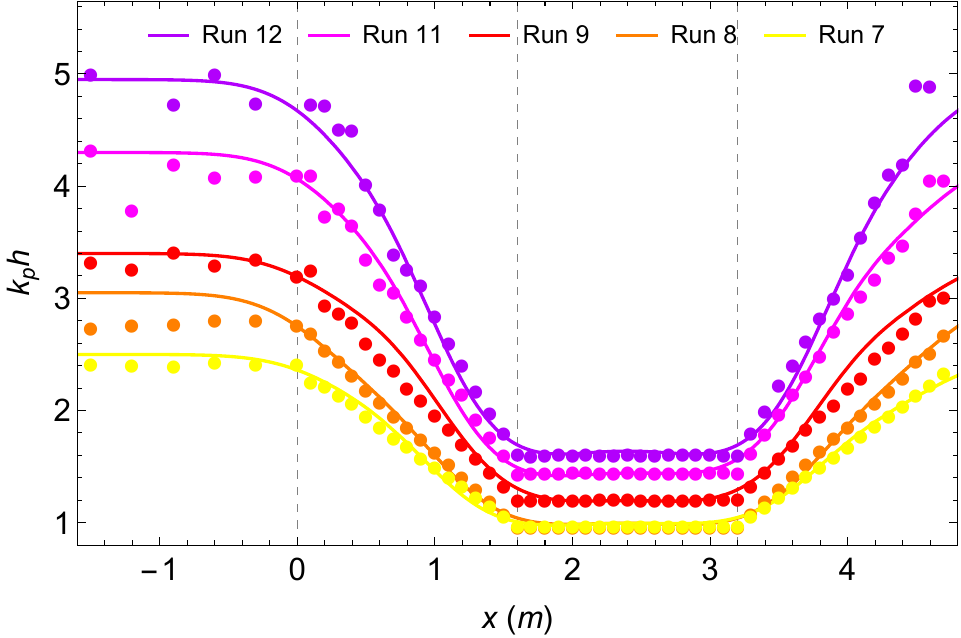}
\endminipage
\caption{Transformation of relative water depth $\xx{k_{p}h}$ over a breakwater according to observation in \citet{Raustol2014} (dots) and numerical fit thereof (solid lines) from \citet{Mendes2021b}. Shoaling ($0 \leqslant x \leqslant 1.6$) and de-shoaling zones ($3.2 \leqslant x \leqslant 4.8$) are marked with dashed vertical lines.}
\label{fig:khevolution}
\end{figure*}
\begin{eqnarray}
\hspace{-0.6cm}
\nonumber
F_{0}^{(2)} &=&  F_{0}^{(1)} +  \rho g  \int_{-h}^{0} \left\langle  \delta u \cdot \delta \zeta^{\star}    \right\rangle \,  dz \quad ,
\\
\nonumber
&=& F_{0}^{(1)} + \rho g a^{2} \cdot   \frac{2\omega }{\sinh{(2\Lambda)}} \cdot \frac{1}{\sinh^{6}{\Lambda}} \times 
\\
\nonumber &{}&
\left(  \frac{3ka}{4}  \right)^{2} \int_{-h}^{0} \cosh^{2}{(2\theta )} \langle \cos^{2}{(2\phi)} \rangle dz  \quad ,
\\
\nonumber
&=&  E^{(1)} \cdot c_{g} \Bigg[  1 +  \left(  \frac{3ka}{4}  \right)^{2}  \frac{1}{\sinh^{6}{\Lambda}} \times
\\
&{}& \left(  1 + \frac{ \sinh{(4\Lambda )} - 2\sinh{(2\Lambda )} }{ 4 \Lambda + 2 \sinh{(2\Lambda )} } \right)
\Bigg] \neq E^{(2)}c_{g} \, .
\label{eq:flux11}
\end{eqnarray}
Hence, the energy flux can not be understood as the product of the group velocity and the exact energy at each higher-order \xx{perturbation} in steepness. \xx{This is of course a well-known fact, see for instance \citet{Higgins1975b} for a general analysis of finite amplitude waves. As pointed out by \citet{Higgins1975b}, the form $F^{(2)} = E^{(2)}c_{g}$ can be recovered only in deep or intermediate waters because the term containing $\sinh^{-6}{\Lambda}$ is very small and the second-order correction to the energy will not exceed the linear term by more than 0.5\%, see eq. 3.10 of \citet{Mendes2021b}. In \jfm{figure} \ref{fig:ratioflux} \xx{the relative growth of} the second-order correction to the energy flux \xx{is compared to the increase in} the energy itself \xx{as computed in} \citet{Mendes2021b}\xx{, illustrating that the energy flux decomposition between energy and group velocity of second-order waves is only possible in deep water.} Furthermore, in wave processes where the system is driven out of equilibrium, the inability to write the energy flux in the same manner of linear theory is even more apparent \citep{Whitham1962}. At higher orders, \citet{Jonsson1995} showed that the energy flux of wave-current interactions can be even written in four different ways, whose form in no way can be traced back to that of the linear theory of free waves.}

\subsection{\x{Shoaling Process over Arbitrary Finite Slopes}}

Now, suppose the water depth change is not adiabatic and derivatives of $h(x)$ can no longer be neglected. The slop-dependent part $\Delta u$ of the horizontal velocity component ($u \rightarrow u + \Delta u$) is written as:
\begin{eqnarray}
\Delta u &=& \frac{a\omega}{k} \Bigg\{ \sin{\phi} \, \pd{}{x} \left[ \frac{\cosh{\theta} }{\sinh{\Lambda}}  \right]  
\\
\nonumber
&{}& + \left(  \frac{3ka}{8}  \right) \sin{(2\phi)} \,  \pd{}{x} \left[ \frac{\cosh{(2\theta)} }{\sinh^{4}{\Lambda}}  \right] \Bigg\} \quad ,
\\
\nonumber
&=&  \frac{a\omega  \nabla h}{\sinh^{2}{\Lambda}}  \left\{ \mathscr{H}_{1} \sin{\phi}    + \left(  \frac{3ka}{4}  \right) \frac{\mathscr{H}_{2} \sin{(2\phi)} }{\sinh^{4}{\Lambda}} \right\} \, ,
\label{eq:U2}    
\end{eqnarray}
with hyperbolic coefficients:
\begin{eqnarray}
\nonumber
\mathscr{H}_{1} &=&  \sinh{\theta} \sinh{\Lambda} - \cosh{\theta} \cosh{\Lambda} \,\, ,
\\
\mathscr{H}_{2} &=&  \sinh{(2\theta)} \sinh^{2}{\Lambda} - \cosh{(2\theta)} \cosh{(2\Lambda)} \,\, .
\end{eqnarray}
The correspondent energy flux is found:
\begin{eqnarray}
F_{0 \, ,\, \nabla h}^{(2)} &=& \rho g  \int_{-h}^{0} \Big\langle \zeta^{\star} \cdot (u + \Delta u)  \Big\rangle \,  dz 
\\
\nonumber
&+&  \rho g  \int_{-h}^{0} \left\langle \frac{(u + \Delta u)}{2g}  \Big[  (u + \Delta u)^{2} + w^{2}   \Big] \right\rangle \,  dz \quad .
\end{eqnarray}
All the pure velocity terms vanish \zz{because the contributions of $
\langle \sin^{2n+1}{\phi} \, \cos^{2m+1}{\phi} \rangle$,  $\langle \sin^{2n}{\phi} \, \cos^{2m+1}{\phi} \rangle $, and $ \langle \sin^{2n+1}{\phi} \, \cos^{2m}{\phi} \rangle )$ vanish for all $(m,n) \in \mathbb{N}^{\ast}$}. The remaining term $\langle \zeta^{\star} \Delta u  \rangle$ has as integrand $\sin{(m\phi)}\cos{(n\phi)}$ and also vanish. Then, \zz{I} find that $F_{0 \, ,\, \nabla h}^{(2)} = F_{0}^{(2)}$ and no slope dependence can be extracted from the energy flux by means of its definition in eq.~(\ref{eq:energyflux}) with mean energy level as \textit{datum}. In other words, the only term containing the slope dependence is the actual set-down flux, \xx{although more terms have to be added in the most general case \citep{Jonsson1995}}. \xx{This exercise highlights the unfeasibility of seeking a slope-dependent correction to the conservation of the energy flux}. Such a task is \xx{quite challenging, inasmuch as the author is unaware of any work that} proper\xx{ly} formulat\xx{es} the energy flux for a \xx{given arbitrary bathymetry}. \xx{Regardless of the feasibility of this endeavor (energy flux generalization), I will} later \xx{propose} a theoretical shortcut of minimal algebraic work based on extreme wave statistics \xx{to effectively solve the} problem \xx{of nonlinear wave transformation over arbitrary bathymetry}. 

\subsection{Observed Shoaling of Irregular Waves \\ over Breakwaters}

\begin{figure*}
\centering
    \includegraphics[scale=0.56]{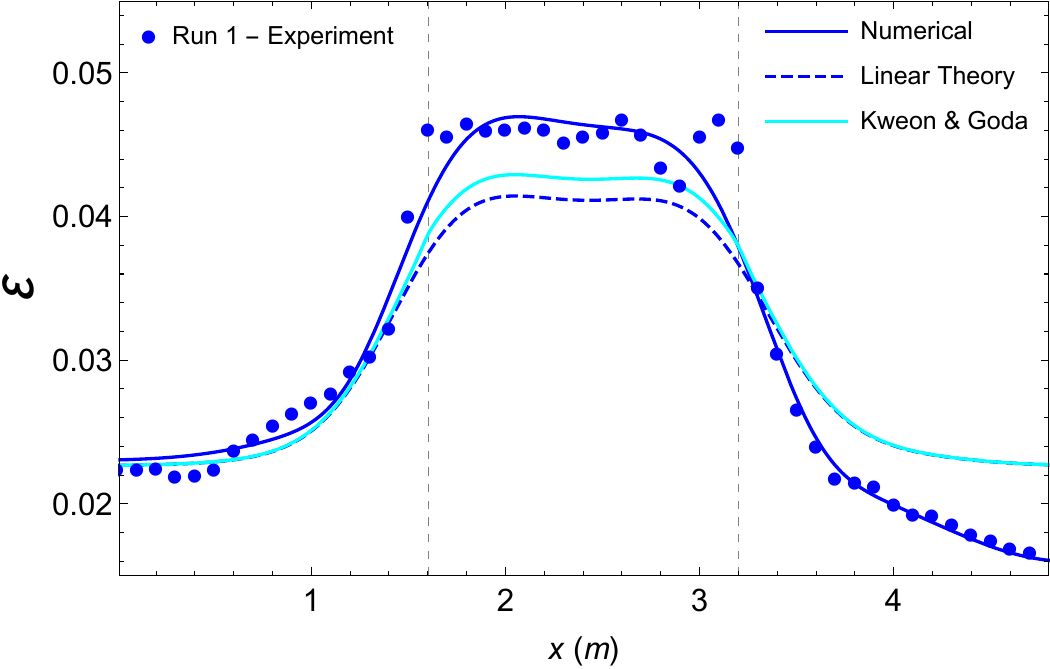}
\caption{Evolution of \xx{the mean irregular wave} steepness \xx{according to linear theory (dashed curve)} compared with observations (dots) \citep{Raustol2014} and its numerical fit (solid curve) \citep{Mendes2021b}.}
\label{fig:steepshoaling}
\end{figure*}
When translating \xx{features of} regular to \xx{those of} irregular waves, the shoaling coefficients of significant wave heights and wavelengths are a good approximation for the regular counterpart \citep{Goda1975}. Hence, one must use \xx{the "significant" wave number (of the 1/3 tallest waves)} $k_{1/3} = 2\pi / \lambda_{1/3}$ within eq.~(\ref{eq:varepsilons2}). Considering empirical relations between peak, energy and 1/3 periods for a broad-banded JONSWAP spectrum \citep{Ahn2021,Mendes2021a}, transformation to peak wavelengths typically follow $ \lambda_{p}/\lambda_{1/3} \sim 1.2 $. \xx{Only then a proper} compar\xx{ison between} the linear theory \xx{for} irregular wave shoaling with \xx{the observations} \xx{can be performed}. \xx{The experiments in} \citet{Raustol2014} \xx{provide a wide range of pre-shoal sea states travelling past a symmetrical breakwater. In} \jfm{figure} \ref{fig:khevolution} \xx{its main results on the evolution of relative water depth are presented. A table containing all physical details of these experimental runs can be found in \citet{Trulsen2020}.} 

\xx{Central to the problem discussed in this work is the evolution of steepness over the breakwater, see for instance \jfm{figure}} \ref{fig:steepshoaling}. The steepness along the shoaling zone is \xx{for the most part} slightly underpredicted by linear theory, but this deviation grows as $k_{p}h$ is lowered \xx{in the vicinity of and at the plateau following} the shoal. Moreover, linear theory strongly overpredicts the steepness \xx{for most of} the de-shoaling zone. \xx{As reviewed in the previous sections, no theoretical or empirical formulae is presently able to predict the nonlinear wave transformation for steep slopes. More importantly, there is also no physical explanation in terms of energy flux conservation for the drop in steepness as compared to linear wave theory in the de-shoaling zone. Nonetheless, since \citeauthor{Shuto1974}'s shoaling theory is acclaimed as the most accurate model for mild slopes \citep{Gupta2017,Ratta2018}, it is good to analyze its predictions for the sake of gaining insight. Using} laboratory data of \xx{mild} slopes, \citet{Goda1997} reformulated the piecewise theory of \citet{Shuto1974} to obtain the relation:
\begin{equation}
K_{\varepsilon , G}^{\ast} = K_{\varepsilon} + \frac{3 \zz{ \coth{(k_{1/3}h)} } }{2000} \left(  \frac{\lambda_{1/3 , 0}}{h} \right)^{2.87} \left(  \frac{H_{s0}}{\lambda_{1/3 , 0}} \right)^{1.27} \quad ,
\label{eq:Godaest}
\end{equation}
where $K_{\varepsilon , G}^{\ast}$ denotes the nonlinear shoaling coefficient\xx{, $H_{s0}$ the significant wave height and $\lambda_{1/3 , 0}$ the significant wavelength in deep water. Converting the wavelength to the spectral peak counterpart \citep{Ahn2021}, two new forms can be found (dimensionless and otherwise):}
\begin{eqnarray}
\nonumber
\xx{
K_{\varepsilon , G}^{\ast} - K_{\varepsilon} } &\approx& \frac{\zz{ \coth{(k_{p}h)} }}{438} \frac{ H_{s0}^{1.27} T_{p0}^{3.2} }{h^{2.87}} 
\\
&\approx&   \frac{\pi \, \zz{ \coth{(k_{p}h)} }}{50\sqrt{2}} \frac{ \varepsilon_{0} }{ (k_{p0}h_{0})^{3} } \left(  \frac{ h_{0}}{ h}  \right)^{3}  \, ,
\label{eq:Godaest2}
\end{eqnarray}
\xx{where $\varepsilon = (\sqrt{2}/\pi) k_{p}H_{s}$ is the steepness measure originated in the non-homogeneous spectral theory of \citet{Mendes2021b,Mendes2022c}. While the observed nonlinear shoaling coefficient in the experiments of \citet{Raustol2014,Trulsen2020} lies in the range $1.07 \leqslant K_{\varepsilon}^{\ast} \leqslant 2$, the correction to linear theory according to eq.~(\ref{eq:Godaest2}) is much smaller} $(K_{\varepsilon , G}^{\ast} - K_{\varepsilon} \lesssim 0.03)$ \xx{due to relatively deep waters prior to the shoal $(k_{p0}h_{0} \geqslant 1.8)$}. \xx{As seen in \jfm{figure} \ref{fig:steepshoaling}, the correction predicted by \citet{Goda1997}, and by extension \citeauthor{Shuto1974}'s theory,} can not describe the \xx{observed} departure from linear theory \xx{over a breakwater. Furthermore, it provides no qualitative improvement over the de-shoaling zone, although the model was not originally formulated for this region. In the next section, I demonstrate how to compute the nonlinear shoaling coefficient compatible with observations, effective for both the maximum atop the shoal and de-shoaling zones}.

\section{Stochastic Formulation for the Wave Transformation}

As unveiled in the previous section \xx{and reviewed in the literature \citep{Gupta2017}}, the best available shoaling theory \citep{Shuto1974} and its empirical parameterization \citep{Goda1997} stumble on the regime of steep slopes \xx{under a wide} range of pre-shoal conditions. This deficiency might stem from the manner it is used to describe the conservation of energy flux itself, and not necessarily the water wave solution. Be that as it may, it is evident that no available theoretical or approximated closed-form shoaling coefficient \xx{is able} to portray the observed laboratory \xx{experiments of steep beaches and breakwaters}. In this section, I attempt to relate statistical tools for the description of irregular waves with its nonlinear slope-dependent shoaling coefficient. Since \citet{Mendes2022b} parameterized the effect \xx{of the slope magnitude} on the probability of rogue waves travelling over a \xx{breakwater}, I shall solve the inverse problem of describing this slope-dependent departure from non-Gaussian seas as a departure in the shoaling coefficient from its linear theory expectation. 

\subsection{Non-homogeneous Wave Statistics}

Let me review the main theoretical aspects of inhomogeneous shoaling wave fields belonging to the works of \citet{Mendes2021b} and \citet{Mendes2022b}. \xx{As a remark, the model does not consider the effects of refraction, breaking, and reflection. This is justified by the low amount of reflection and absence of breaking in the experiments of \citet{Trulsen2020}, see \citet{Benoit2021} for a review.} \zz{Perturbative up to} $m$-th order in steepness, the generalized velocity potential $\Phi$ and surface elevation $\zeta$ solutions are given:
\begin{eqnarray}
\Phi =  \sum_{m} \Omega_{m} \frac{\cosh{(m\varphi)} \sin{(m \phi)}}{mk} \,\,  ;  \,\,
\zeta = \sum_{m} \Tilde{\Omega}_{m} \cos{(m\phi)} \,\, .
\label{eq:potentialsurface}
\end{eqnarray}
For waves of second-order in steepness one has:
\begin{eqnarray}
\nonumber
\Omega_{1} &=& \frac{a\omega}{ \sinh{kh}} \quad ; \quad \Omega_{2} =  \frac{3ka^{2} \omega}{ 4 \sinh^{4}{kh}} \,\, ; 
\\
\Tilde{\Omega}_{1} &=& a \quad ; \quad \Tilde{\Omega}_{2} = \frac{ka^{2} }{ 4 } \left[ \frac{3 - \tanh^{2}{(kh)} }{ \tanh^{3}{(kh)}  }  \right]  \, .
\label{eq:D7}
\end{eqnarray} When this group of waves is affected by a change in bathymetry, the \citet{Khinchin1934} theorem \xx{is modified such that the} follow\xx{ing ratio no longer equals unity}:
\begin{eqnarray}
\Gamma (x) := \frac{\mathbb{E}[\zeta^{2}] }{ \mathscr{E} }   = \frac{\mathbb{E}[\zeta^{2}(x , t)](x) }{ \mathscr{E}(x)   }  \quad ,
\label{eq:GammaEnsemble}
\end{eqnarray}
where $\mathbb{E}[\zeta^{2}]$ is the ensemble average of the square of the surface elevation and $\mathscr{E}(x)$ the spectral energy density, i.e. gravity and water density are factored out. The energy density of small amplitude waves travelling over a steep slope satisfying $ |\nabla h| \geqslant 1/20$ up to second order in steepness $(m \leqslant 2)$ reads:
\begin{eqnarray}
\nonumber
\mathscr{E} &=& \sum_{m} \frac{\tilde{\Omega}_{m}^{2}}{4} + \sum_{m} \frac{\Omega_{m}^{2}}{4g} \int_{-h}^{0} \cosh{\big(2m\varphi \big)}  \, dz  \,\, , 
\\
\nonumber
&=& \frac{1}{4} \sum_{m} \left[ \tilde{\Omega}_{m}^{2} + \Omega_{m}^{2} \cdot \frac{ \sinh{(2mk_{p}h)}}{2mgk_{p}} \right] \, ,
\\
 &=& \sum_{i} \frac{a_{i}^{2}}{2} \left[  1+  \frac{\pi^{2} \varepsilon^{2}\mathfrak{S}^{2}  }{16} \left(  \frac{\Tilde{\chi}_{1} + \chi_{1}}{2}  \right) \right] \,\, ,  
\label{eq:energydef}
\end{eqnarray}
where $\mathfrak{S} = ka / \pi \varepsilon$ is the wave vertical asymmetry \xx{measuring the mean ratio between wave crest and wave heights}, and with trigonometric coefficients:
\begin{equation}
\,\, \Tilde{\chi}_{1} =  \left[ \frac{3 - \tanh^{2}{(k_{p}h)} }{ \tanh^{3}{(k_{p}h)}  } \right]^{2}  \,\, ; \,\, \chi_{1} = \frac{9\, \textrm{cosh}(2k_{p}h) }{\textrm{sinh}^{6} (k_{p}h)} \quad .
\end{equation}
On the other hand, assuming a weakly stationary process of waves travelling over a shoal while inhomogeneous in space, one may approximate the ensemble average by the time average and the variance of the surface elevation now reads:
\begin{eqnarray}
\nonumber
\langle \zeta^{2} \rangle_{t} &=&  \frac{1}{T}  \int_{0}^{T} \left[ \sum_{m}  \tilde{\Omega}_{m} \, \cos{(m\phi)} \right]\left[ \sum_{n}  \tilde{\Omega}_{n} \, \cos{(n\phi)} \right] dt  ,
\\
 &=& \sum_{m} \frac{\tilde{\Omega}_{m}^{2}}{2}  =  \sum_{i} \frac{a_{i}^{2}}{2} \left[  1+  \frac{\pi^{2} \varepsilon^{2}\mathfrak{S}^{2}  }{16}  \Tilde{\chi}_{1}  \right] \quad .
\label{eq:B6}
\end{eqnarray}
If waves are linear ($\varepsilon \rightarrow 0$), the process is ergodic, stationary and homogeneous by means of $\mathbb{E}[\zeta^{2}] = \langle \zeta^{2} \rangle_{t} = \mathscr{E} = \sum_{i} a_{i}^{2}/2$ and \xx{thus} $\Gamma = 1$. Then, it can be shown that the narrow-banded distribution of wave heights up to second order in steepness will be transformed to account for $\Gamma$:
\begin{eqnarray}
\hspace{0.0cm}
\mathcal{R}_{\alpha,\Gamma}(H>\alpha H_{s}) = e^{-2\alpha^{2}/\Gamma} \quad ,
\label{eq:Rayexc}
\end{eqnarray}
with the analytical expression for the second-order inhomogeneity correction $\Gamma$ reading:
\begin{eqnarray}
\Gamma  =\frac{ 1+   \frac{\pi^{2} \varepsilon^{2} \mathfrak{S}^{2}}{16}     \, \Tilde{\chi}_{1} }{  1+   \frac{\pi^{2} \varepsilon^{2} \mathfrak{S}^{2}}{32}   \, \left( \Tilde{\chi}_{1}  + \chi_{1} \right)  } \quad .
\label{eq:Gamma}
\end{eqnarray}
\xx{Recently, this theory has been successfully} generalized \xx{to an arbitrary bathymetry shape of mean} slope \xx{magnitude} $|\nabla h |$ \citep{Mendes2022b,Mendes2023b}:
\begin{eqnarray}
\Gamma_{\nabla h} =  \frac{ 1+   \frac{\pi^{2}\mathfrak{S}^{2} \varepsilon^{2}}{16}      \, \Tilde{\chi}_{1} }{  1+   \frac{\pi^{2} \mathfrak{S}^{2} \varepsilon^{2}}{32}   \, \left( \Tilde{\chi}_{1}  + \chi_{1} \right) + \check{\mathscr{E}}_{p2} }           \,\, ,
\label{eq:slopegamma0}      
\end{eqnarray}
where $\check{\mathscr{E}}_{p2}$ is the change in potential energy due to a finite mean water level $\langle \zeta \rangle \neq 0$ under the presence of waves as compared to the still water level:
\begin{equation}
\check{\mathscr{E}}_{p2} = \frac{6\pi^{2}\varepsilon^{2}}{5\mathfrak{S}^{4}(k_{p}h)^{2}}   \, \tilde{\nabla} h \Big(1 + \tilde{\nabla} h\Big) \,\, , \,\,   \tilde{\nabla} h \equiv \frac{\pi \nabla h}{k_{p0}h_{0}} \, .
\label{eq:Ep2}
\end{equation}

\subsection{Inversion of the Slope Effect}

\begin{figure*}
\centering
    \includegraphics[scale=0.6]{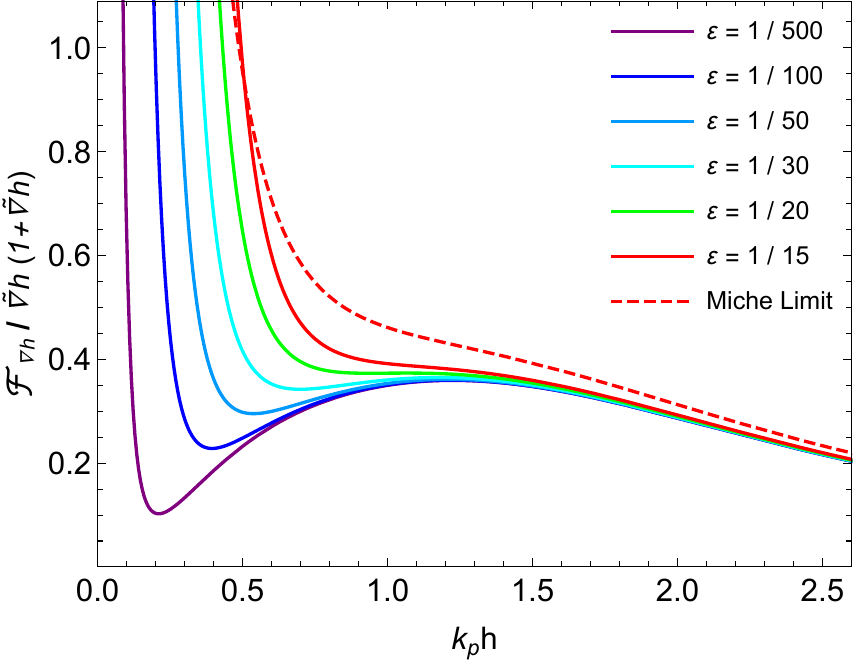}
\caption{Evolution of \zz{the sea state parameters} $\mathcal{F}_{\nabla h} / \tilde{\nabla} h \Big(1 + \tilde{\nabla} h\Big)$ with initial relative water depth $k_{p}h_{0}=\pi$ and asymmetry $\mathfrak{S} \approx 1.2$ \xx{\citep{Mendes2022c}} according to eq.~(\ref{eq:slopecorrnonlin2}).}
\label{fig:trigonom}
\end{figure*}
\begin{figure*}
\minipage{0.5\textwidth}
    \includegraphics[scale=0.58]{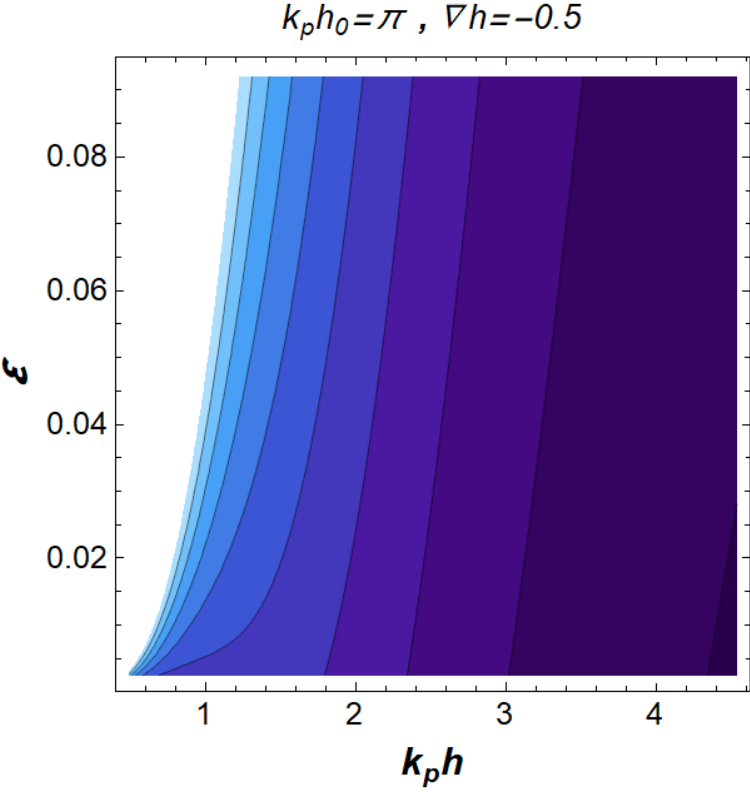}
\endminipage
\hfill
\minipage{0.49\textwidth}
    \includegraphics[scale=0.58]{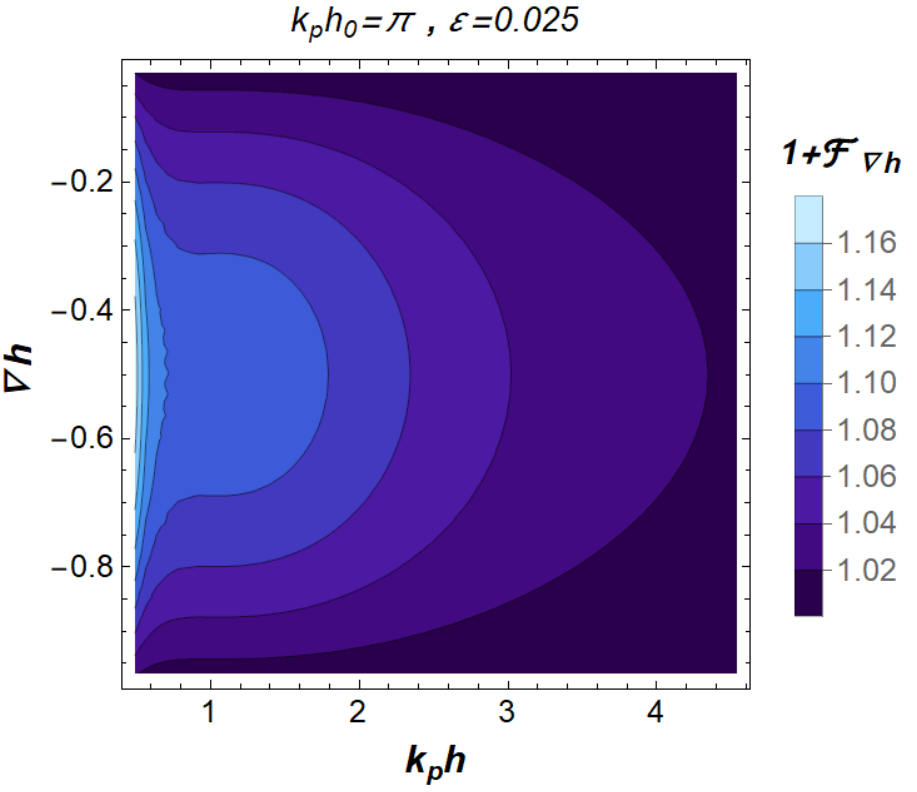}
\endminipage

\vspace{0.5cm}
\minipage{0.52\textwidth}
    \includegraphics[scale=0.56]{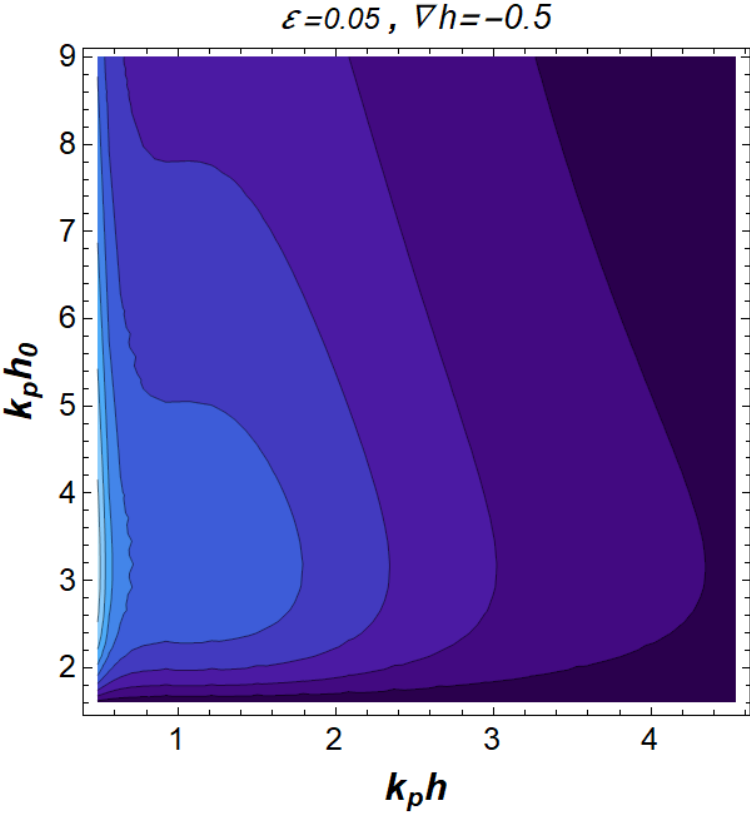}
\endminipage
\hfill
\minipage{0.48\textwidth}
    \includegraphics[scale=0.56]{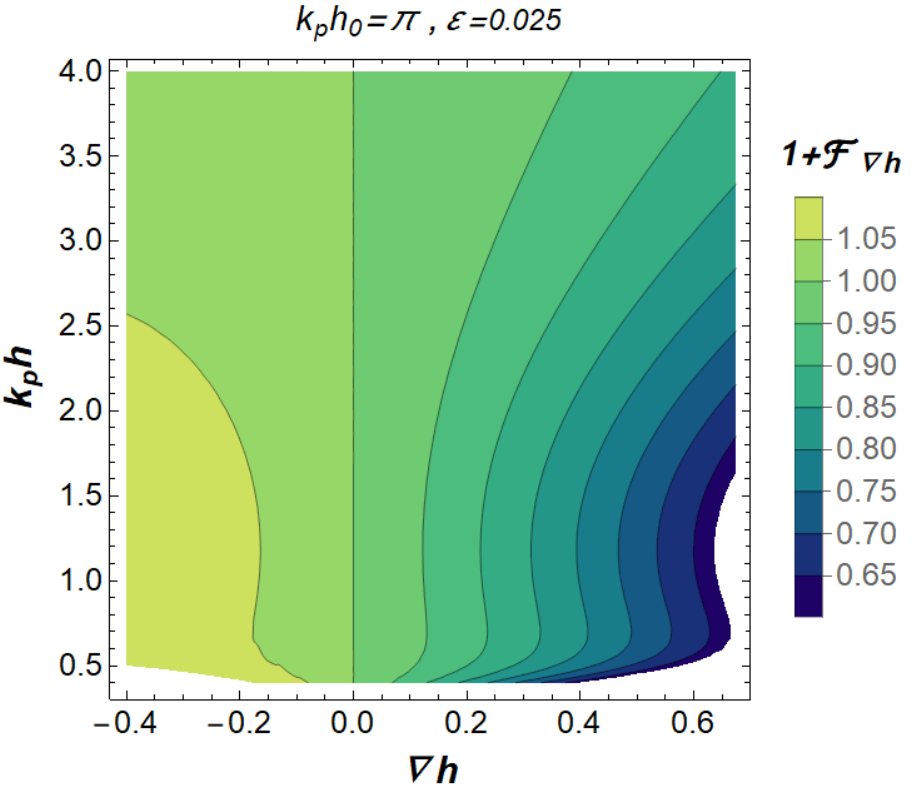}
\endminipage
\caption{Nonlinear correction to the shoaling coefficient as a function of relative depth, pre-shoal relative depth, mean wave steepness and slope magnitude of the shoal/de-shoal.}
\label{fig:contournonl}
\end{figure*}
\xx{The maximum drop in the mean water level due to shoaling appears near the top of the breakwater and quickly vanishes further atop \citep{Gourlay2000}. In the case of small amplitude waves and absence of wave breaking, the decrease of the set-down will still be manifested atop the shoal of a breakwater, because the latter induces a piling up in the de-shoaling zone \citep{calabrese20082d}. As such, the effect of $\check{\mathscr{E}}_{p2}$ quickly grows near the peak of mean water level drop in the shoaling zone while the wave transformation can still be well described by linear theory. On the other hand, with fractions of wavelength atop the shoal the nonlinear wave transformation kicks in when the mean water level quickly returns to the pre-shoal level. From a statistical perspective, the increase in exceedance probability encoded in $\check{\mathscr{E}}_{p2}$ due to the mean water level can be understood as being carried out and transferred to the nonlinear shoaling of the wave steepness.} Suppose I may write $K^{\ast}_{\varepsilon} = K_{\varepsilon} (1+\mathcal{F}_{\nabla h})$ for the nonlinear modification of the linear shoaling coefficient. Then, the non-homogenous spectral correction $\Gamma$ depends on the slope under the regime of linear shoaling over the ramp and is equivalent to the regime of nonlinear shoaling \xx{atop the breakwater}:
\begin{eqnarray}
\Gamma_{\nabla h} \approx \frac{ 1+   \frac{\pi^{2}}{16}  \mathfrak{S}^{2} \varepsilon^{2} (1+\mathcal{F}_{\nabla h})^{2}   \, \Tilde{\chi}_{1} }{  1+   \frac{\pi^{2} }{32} \mathfrak{S}^{2} \varepsilon^{2} (1+\mathcal{F}_{\nabla h})^{2}  \, \left( \Tilde{\chi}_{1}  + \chi_{1} \right)  }     \, .
\label{eq:slopegamma}
\end{eqnarray}
Let the definition $\Omega^{2}_{\varepsilon} = (\pi^{2}/16) \, \mathfrak{S}^{2} \varepsilon^{2}$ be employed. Hence, one finds the modification to the nonlinear shoaling coefficient through:
\begin{eqnarray}
\frac{ 1+   \Omega^{2}_{\varepsilon}  \Tilde{\chi}_{1} }{      1+  \Omega^{2}_{\varepsilon} \frac{ \left( \Tilde{\chi}_{1}  + \chi_{1} \right) }{2}  + \check{\mathscr{E}}_{p2}} \approx  \frac{    1+   \Omega^{2}_{\varepsilon} (1+\mathcal{F}_{\nabla h})^{2}  \Tilde{\chi}_{1}   }{   1+  \Omega^{2}_{\varepsilon} (1+\mathcal{F}_{\nabla h})^{2}  \frac{\left( \Tilde{\chi}_{1}  + \chi_{1} \right)}{2} } 
 \quad .
\end{eqnarray}
Solving the equation for $\mathcal{F}_{\nabla h}$, I obtain:
\begin{equation}
\mathcal{F}_{\nabla h} \approx \sqrt{ \frac{ \Omega^{2}_{\varepsilon}  \left( \Tilde{\chi}_{1}  - \chi_{1} \right)  - 2\check{\mathscr{E}}_{p2} }{ \Omega^{2}_{\varepsilon}  \left( \Tilde{\chi}_{1}  - \chi_{1} \right)  + 2\check{\mathscr{E}}_{p2} \Omega^{2}_{\varepsilon} \Tilde{\chi}_{1}}   } - 1 \quad .
\end{equation}
Since $\check{\mathscr{E}}_{p2}$ represents only a small fraction of either potential or kinetic spectral energies, a Taylor expansion up to first order in $\check{\mathscr{E}}_{p2}/\left( \Tilde{\chi}_{1}  - \chi_{1} \right)$ may be performed, finding:
\begin{eqnarray}
\nonumber
\mathcal{F}_{\nabla h} &=& \sqrt{ 1 - \frac{  2\check{\mathscr{E}}_{p2} ( 1 +  \Omega^{2}_{\varepsilon} \Tilde{\chi}_{1}  ) }{ \Omega^{2}_{\varepsilon}  \left( \Tilde{\chi}_{1}  - \chi_{1} \right)}   } - 1 \,\, ,
\\
&\approx& - \frac{  \check{\mathscr{E}}_{p2} ( 1 +  \Omega^{2}_{\varepsilon} \Tilde{\chi}_{1}  )  }{ \Omega^{2}_{\varepsilon}  \left( \Tilde{\chi}_{1}  - \chi_{1} \right)}    \quad .
\label{eq:slopecorrnonlin}
\end{eqnarray} 
Qualitatively, the slope-dependence of the nonlinear shoaling coefficient is proportional to the normalized variance of the surface elevation \xx{and} the change in potential energy (due to the set-down)\xx{, while being inversely proportional to} the \xx{difference between second-order harmonic corrections. Within the second-order approximation, the difference $\Tilde{\chi}_{1}  - \chi_{1}$ is always finite. These two harmonic corrections are bound to be equal in the limit of $k_{p}h \rightarrow 0$, but in this limit, finite-amplitude wave theory has to be taken into account, precluding this difference from vanishing. Therefore, although the slope-dependent correction to linear shoaling theory will quickly grow in shallower water, it does not diverge if I no longer assume waves of small amplitude (see \jfm{figure} \ref{fig:trigonom}). Indeed}, the model is out \xx{its} scope for $k_{p}h \leqslant (3\pi \varepsilon)^{1/3}$ from the limitations \xx{upon} the Ursell number \xx{by the} second-order theory \citep{Dalrymple984}. As such, this divergence will be balanced \xx{by} wave dissipation. Slope magnitude aside, if waves are linear their mean steepness is small and therefore the normalized variance of the surface elevation recovers unity, both set-down and second-order correction to the energy are negligible. Under these conditions, the linear shoaling coefficient is an accurate portrayal of wave transformation in deep water \xx{regardless of how steep waves are}, see \jfm{figure} \ref{fig:trigonom}. \xx{Likewise, if the wave steepness is very low the linear shoaling will withstand even near shallow water}.  The full formula for the nonlinear shoaling coefficient reads:
\begin{equation}
\frac{ K_{\varepsilon}^{\ast} }{ K_{\varepsilon} } \approx   1 - \frac{  96   \left( 1 +  \frac{\pi^{2}\mathfrak{S}^{2} \varepsilon^{2}}{16}  \Tilde{\chi}_{1}  \right) }{ 5 \mathfrak{S}^{6} (k_{p}h)^{2} \left( \Tilde{\chi}_{1}  - \chi_{1} \right)} \cdot \frac{\pi \nabla h}{ k_{p0}h_{0} } \left( 1 + \frac{\pi \nabla h}{ k_{p0}h_{0} }  \right)   \, .
\label{eq:slopecorrnonlin2}
\end{equation}   
In \jfm{figure} \ref{fig:contournonl} contour plots \xx{are given and describe} how the nonlinear correction to the shoaling coefficient varies with mean wave steepness, \xx{depth gradient for both shoaling and de-shoaling zones, as well as} relative water depth. \xx{The more nonlinear waves are prior to the shoal the larger deviation from linear shoaling will be at steep slopes. While continuous shoaling will lead to dissipation till it reaches a saturation of nonlinear shoaling, the saturation of the effect of slope in the shoaling zone \citep{Mendes2022b} will be translated in a saturation of the increase of the nonlinear shoaling coefficient even without dissipation. For the de-shoaling zone, the decrease in shoaling coefficient is associated with a higher mean water level and is not expected to saturate with steeper slopes.}
\begin{figure*}
\minipage{0.5\textwidth}
    \includegraphics[scale=0.52]{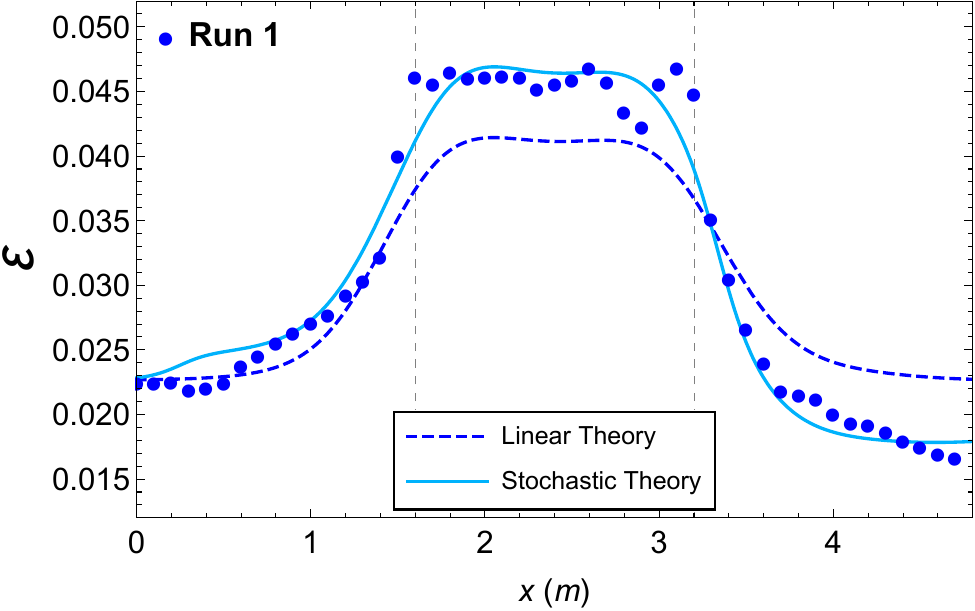}
\endminipage
\hfill
\minipage{0.49\textwidth}
    \includegraphics[scale=0.52]{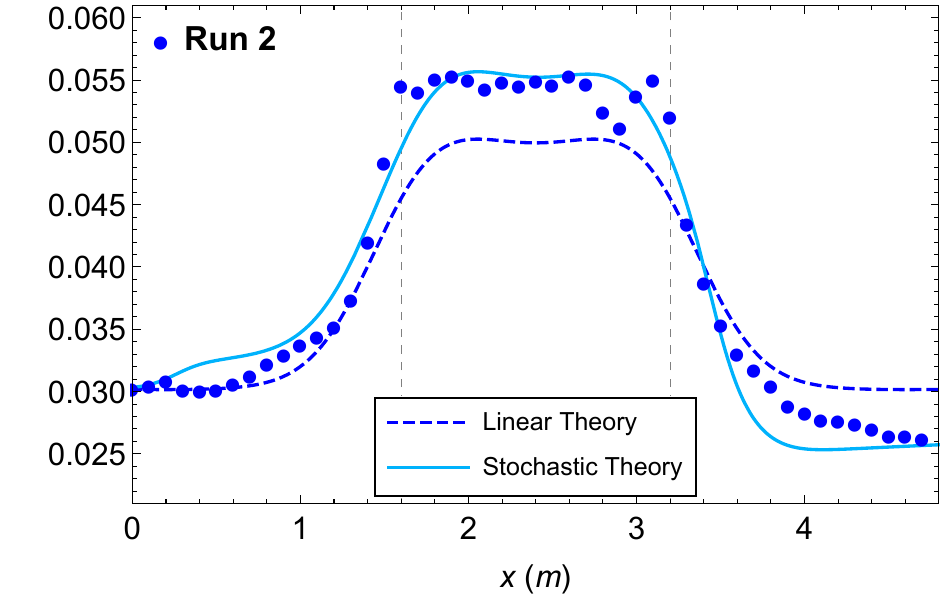}
\endminipage

\minipage{0.5\textwidth}
    \includegraphics[scale=0.52]{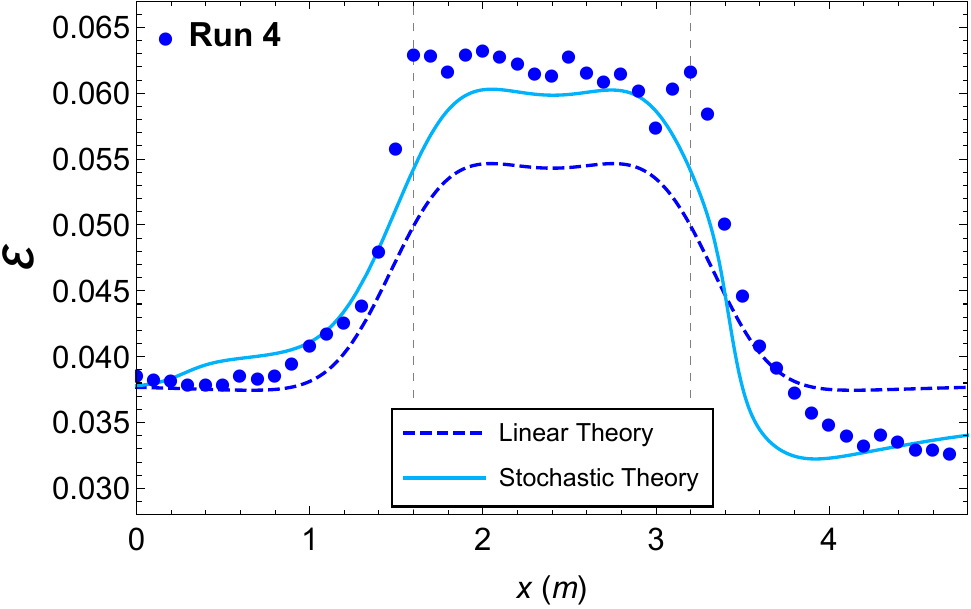}
\endminipage
\hfill
\minipage{0.49\textwidth}
    \includegraphics[scale=0.52]{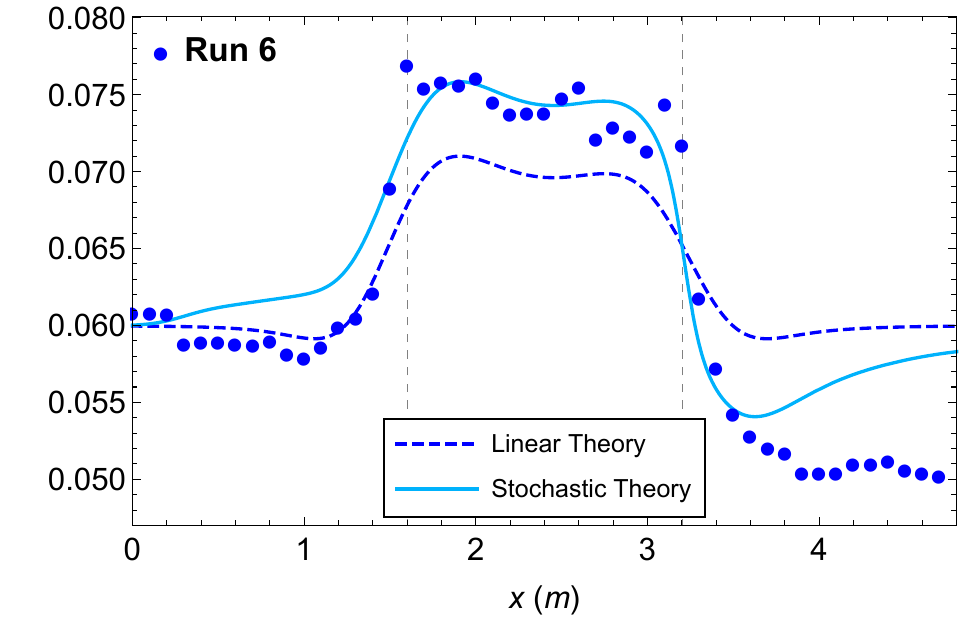}
\endminipage

\minipage{0.5\textwidth}
    \includegraphics[scale=0.52]{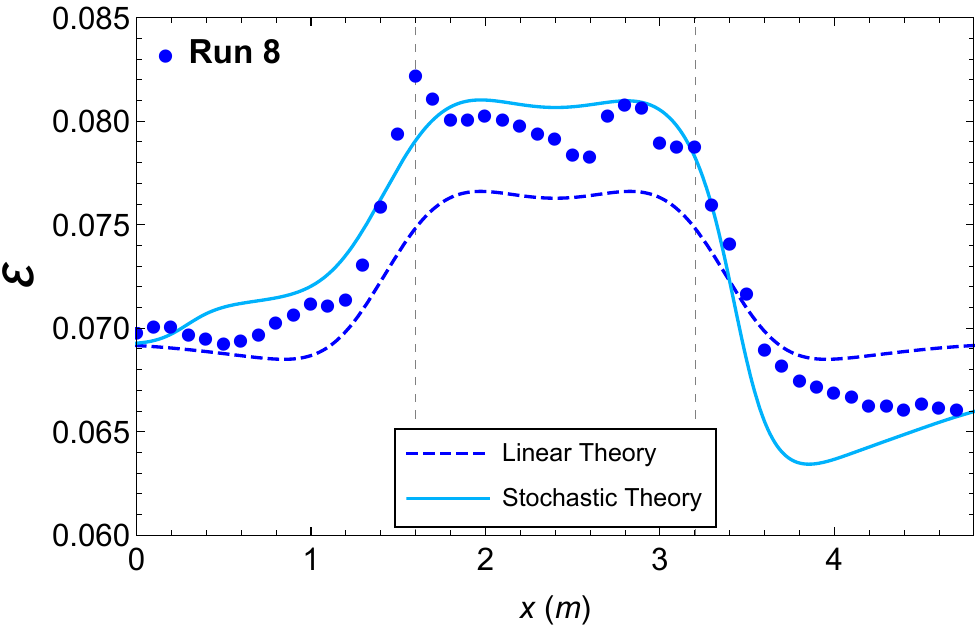}
\endminipage
\hfill
\minipage{0.49\textwidth}
    \includegraphics[scale=0.52]{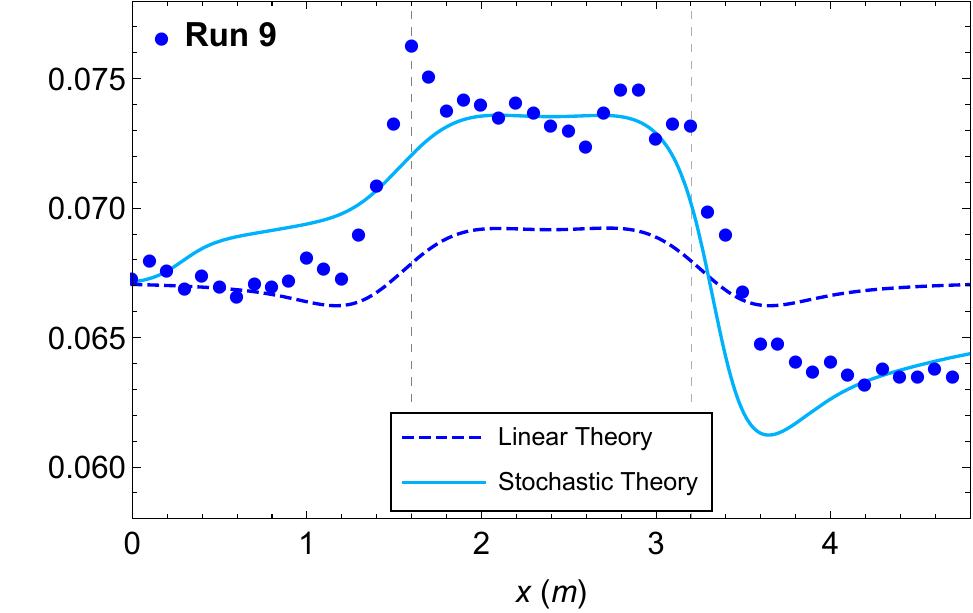}
\endminipage
\caption{New model prediction for the shoaling coefficient of significant steepness measured against observations in \citet{Raustol2014} and its linear theory modelling thereof.}
\label{fig:steepshoaling3}
\end{figure*}
\begin{figure}
\minipage{0.48\textwidth}
    \includegraphics[scale=0.52]{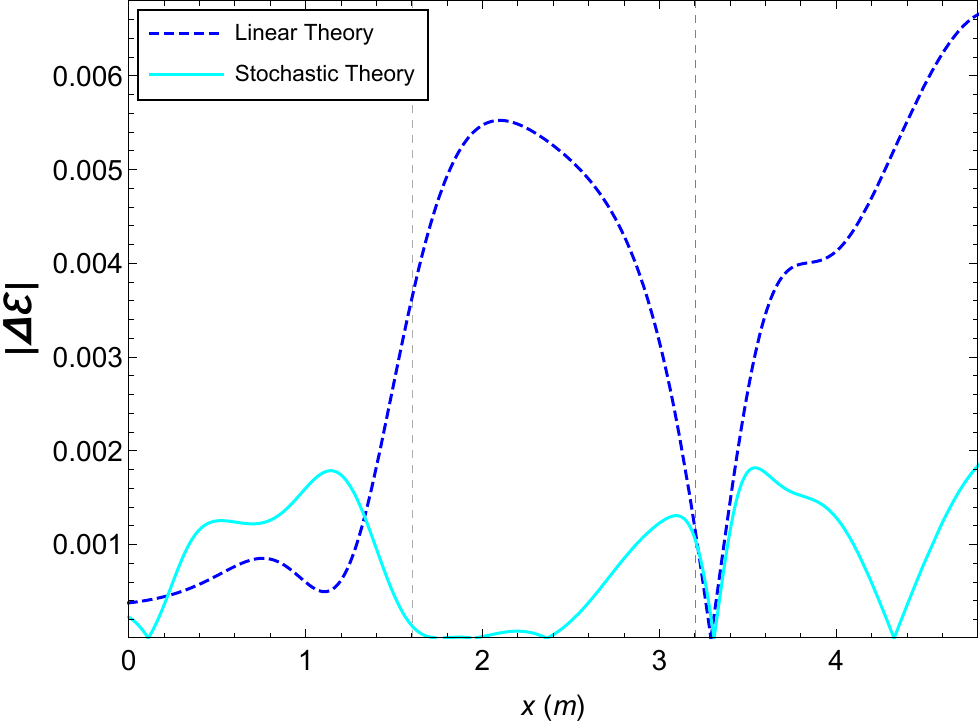}
\endminipage
\hfill
\minipage{0.48\textwidth}
    \includegraphics[scale=0.5]{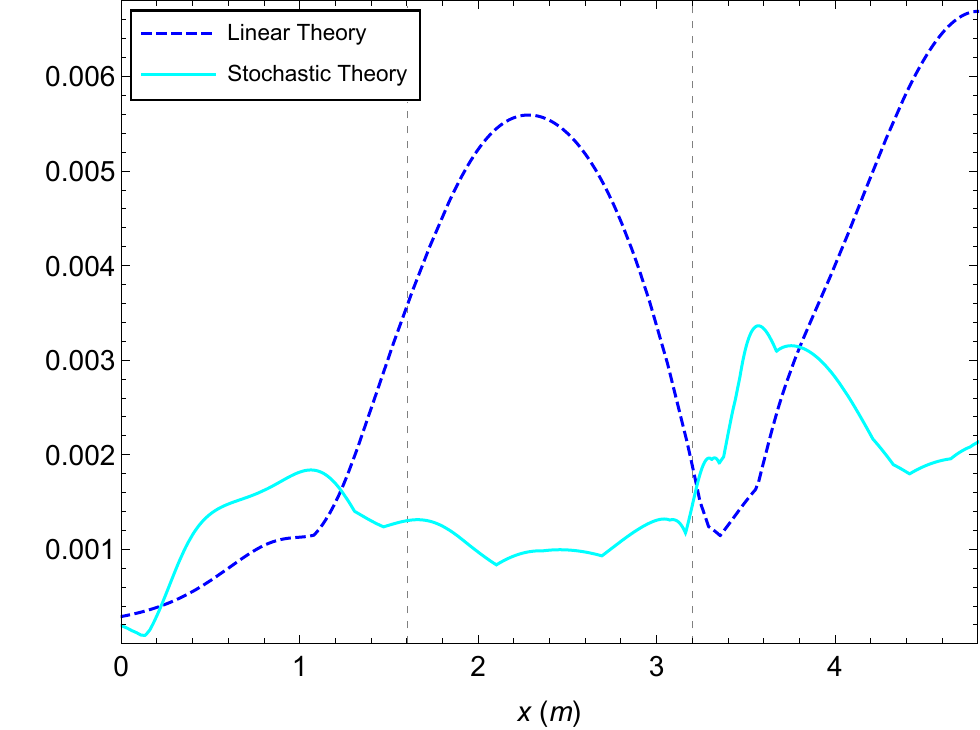}
\endminipage
\caption{Absolute difference between experimental and theoretical steepness.}
\label{fig:steepshoaling4}
\end{figure}

\subsection{Comparison with Experiments}

\xx{Withal,} eq.~(\ref{eq:slopecorrnonlin2}) \xx{is examined against experimental observation of a wide range of initial water wave conditions from \citet{Raustol2014}. These experiments were later summarized in \citet{Trulsen2020} with a focus on the wave statistics, though omitting the evolution of key parameters such as relative water depth and wave steepness.} \jfm{Figure} \ref{fig:steepshoaling3} \xx{explores different runs (panels) organized from shallower to deeper relative depths. The overall trend of nonlinear shoaling is well captured by the present model whereas the linear model is found to be consistently unreliable. However, over about half of the shoaling zone ($0< x <1.2$) the system is weakly nonlinear and the linear and stochastic theories describe the evolution of the steepness equally well. Immediately after this zone, in a transition between shoaling and plateau regions of the breakwater ($1.2< x <1.8$) the mean water level reaches its lowest point (see \citet{Gourlay2000}), and the system is at the strongest departure from linearity. As a result, the stochastic theory is a much better fit for the experiments. This trend is again observed further atop the shoal ($1.8< x <2.8$). Both during the transition between a set-down and a piling-up near the start of the de-shoaling zone ($2.8< x <3.4$) and the remaining part of the de-shoal ($3.4< x <4.8$) the stochastic model provides the best description of the observations, albeit in some cases the magnitude of the decrease in steepness compared to linear theory is larger than in the experimental data. The occasional and small overprediction by the present model of the wave transformation in the shoaling zone and underprediction over the de-shoaling zone is likely due to a seldom overestimation of set-down and piling-up magnitudes. The latter highlights limitations in the slope parameterization of \citet{Mendes2022b}.}

\begin{figure}
\minipage{0.45\textwidth}
    \includegraphics[scale=0.55]{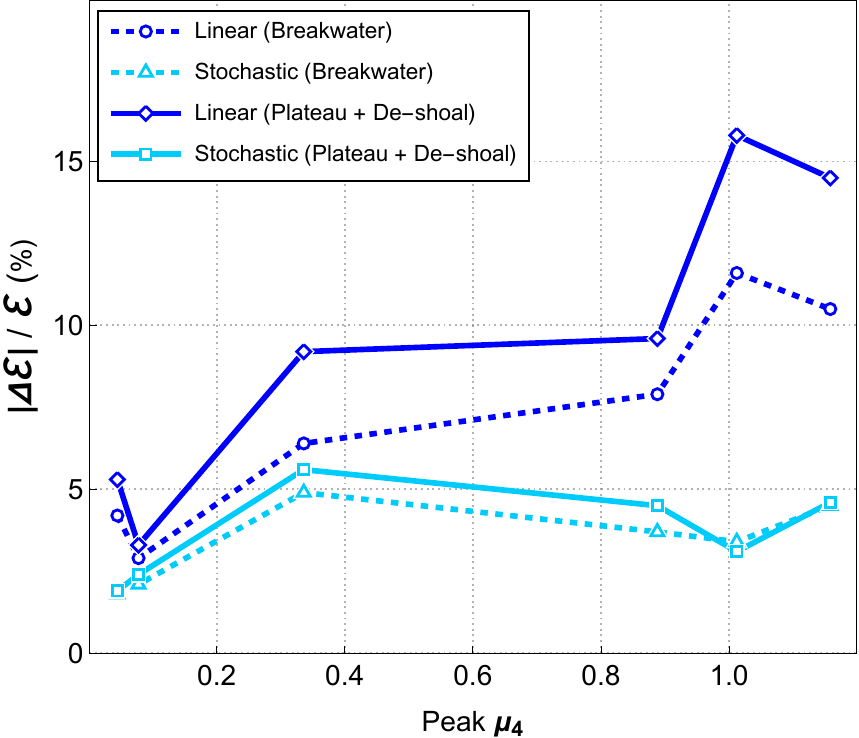}
\endminipage
\hfill
\minipage{0.48\textwidth}
    \includegraphics[scale=0.55]{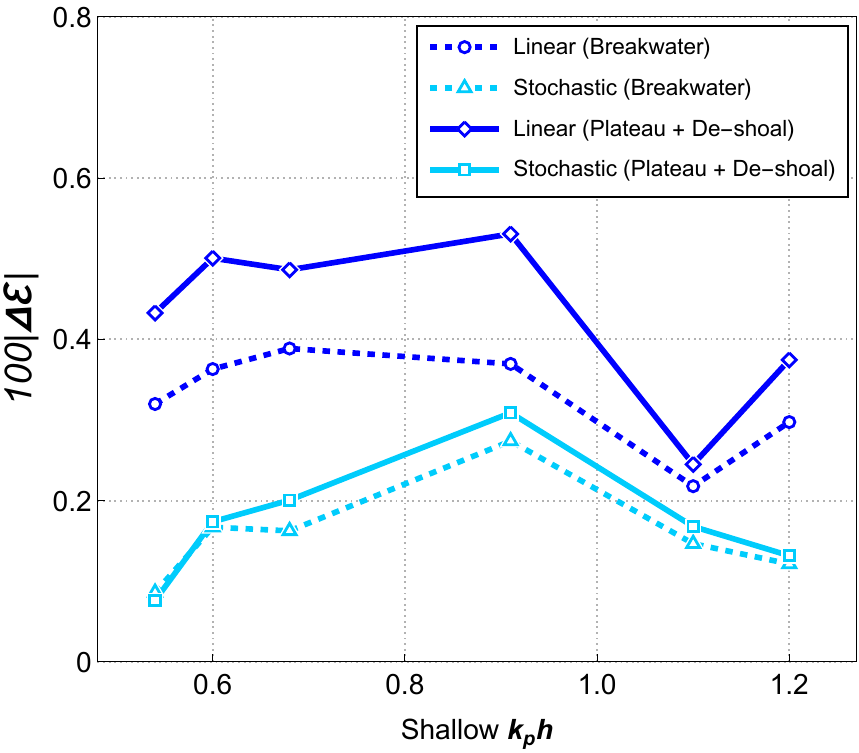}
\endminipage
\caption{Normalized absolute difference between experimental and theoretical steepness as a function of maximum excess kurtosis atop the shoal and the relative water depth.}
\label{fig:steepshoaling5}
\end{figure}

\xx{Further scrutiny in the deviations from linear theory is applied in \jfm{figure} \ref{fig:steepshoaling4} through the evaluation of absolute differences: the stochastic theory outperforms the linear theory atop the shoal and in the de-shoaling zone by almost an order of magnitude, where the nonlinear effects are most pronounced. However, in the region when nonlinear effects are still developing, the two models provide similar deviations from observation. Indeed, the stochastic model has a typical absolute average difference of 3\% from observation through the entire breakwater evolution, whereas the linear theory averages 8\%\xx{. Their difference is not only in magnitude, \jfm{figure} \ref{fig:steepshoaling5} shows that the stochastic model is consistent in all regimes while the linear theory rapidly departs from observations as the peak in excess kurtosis in the breakwater plateau increases. Clearly, this trend is mirrored in the comparison of error as a function of the plateau relative depth $k_{p}h$}. When considering the combined regions of the plateau and de-shoal of the breakwater, the stochastic model deviation increases only slightly and the linear counterpart rises to 11\%. Hence, the comparison throughout multiple regions of the steep breakwater between experiments, stochastic nonlinear and linear theory suggests that the slope magnitude parameterization \citep{Mendes2022b} returns an accurate prediction for the wave transformation, with an ever larger disparity between the models as nonlinear effects grow in intensity in shallower waters.}

Interestingly, a closer look beyond numerical accuracy shows that eq.~(\ref{eq:slopecorrnonlin2}) qualitatively agrees with the estimate from eq.~(\ref{eq:Godaest2}) by assigning no role for the mean steepness in deep water \xx{because $\varepsilon \ll (kh)^{3}$}. This behaviour in deep water must be recovered by any shoaling theory, as deep water waves can not feel the bottom regardless of the slope magnitude. Remarkably, both the stochastic theory in eq.~(\ref{eq:slopecorrnonlin2}) and the numerical fit of \citet{Goda1997} for the nonlinear correction are inversely proportional to $(kh)^{3}$. \xx{This similarity points to a unveiled convergence of the portrayal of the physical system by both deterministic and stochastic approaches.}

\section{Conclusions}

A \xx{stochastic} model for the evolution of steepness in terms of a nonlinear shoaling coefficient has been proven excellent in describing observations of well-known experiments over a steep symmetrical breakwater. \xx{The} formulation proves to be much simpler and more effective than \xx{existing theoretical and empirical models based on} the conservation of energy flux. While \xx{the new} model fills the gap created by the underprediction of the steepness growth of shoaling waves by linear theory, it also describes the \xx{why linear theory overpredicts the} \xx{steepness evolution over} de-shoaling zones. \xx{The main novelty in this study is the approach: the principle of continuity of the stochastic and statistical behavior of the water waves implies a process that can be understood as a transfer from the nonlinearity due to the change in mean water level at the end of the shoaling zone to the shoaling coefficient atop the breakwater. Especially for out-of-equilibrium systems, this symbiosis suggests that it may be possible to compute fundamental properties of nonlinear waves through stochastic means that are otherwise elusive to deterministic methods, such as the conservation of integral properties. Because the new theory deals with small amplitude waves, wave breaking can not be considered. Thus, future work has to address a generalized shoaling coefficient from small up to finite and breaking amplitude waves. Nevertheless, the model can be applied to a beach as well, and the departure of the wave transformation from linear theory will be captured in the region near the peak of set-down.}\\[0.05cm]

\textbf{Declaration of Interests}. The author reports no conflict of interests.

\bibliography{Maintext}% Produces the bibliography via BibTeX.

\end{document}